\documentclass[aps,prb,onecolumn,superscriptaddress,notitlepage]{revtex4-1}
\usepackage{blindtext}
\usepackage{amsmath,amssymb,amsfonts}
\usepackage{color}
\usepackage{stackengine}
\usepackage{epstopdf}
\usepackage[normalem]{ulem} % either use this (simple) or
\usepackage{soul} % use this (many fancier options)

\usepackage{graphicx}
\usepackage{indentfirst}
\usepackage{psfrag}
\usepackage{epsfig}

\newcommand{\QPEa}{\ensuremath{\mathcal{Q}_1^{\textrm{(PE)}}}}
\newcommand{\QPEb}{\ensuremath{\mathcal{Q}_2^{\textrm{(PE)}}}}
\newcommand{\QMB}{\ensuremath{\mathcal{Q}_1^{\textrm{(MB)}}}}
\newcommand{\neff}{\ensuremath{n_\text{eff}}}

\begin{document}

	\title{
		Suspended mid-infrared waveguides for Stimulated Brillouin Scattering
	}
	\author{Miko\l{}aj K. Schmidt}
	\email{mikolaj.schmidt@mq.edu.au}
	\affiliation{Macquarie University Research Centre in Quantum Science and Technology (QSciTech), MQ Photonics Research Centre, Department of Physics and Astronomy, Macquarie University, NSW 2109, Australia.}
	\author{Christopher G. Poulton}
	\affiliation{School of Mathematical and Physical Sciences, University of Technology Sydney, NSW 2007, Australia.}
	\author{Goran Z. Mashanovich}
	\affiliation{Optoelectronics Research Centre, University of Southampton, University Road, Southampton, Hampshire SO17 1BJ, UK.}
	\author{Graham T. Reed}
	\affiliation{Optoelectronics Research Centre, University of Southampton, University Road, Southampton, Hampshire SO17 1BJ, UK.}
	\author{Benjamin J. Eggleton}
	\affiliation{The University of Sydney Nano Institute (Sydney Nano), The University of Sydney, NSW 2006, Australia.}
	\affiliation{ Institute of Photonics and Optical Science (IPOS), School of Physics, The University of Sydney, NSW 2006, Australia.}
	\author{M.J. Steel}
	\affiliation{Macquarie University Research Centre in Quantum Science and Technology (QSciTech), MQ Photonics Research Centre, Department of Physics and Astronomy, Macquarie University, NSW 2109, Australia.}
	
	\begin{abstract}
We theoretically investigate a new class of silicon waveguides for achieving Stimulated Brillouin Scattering (SBS) in the mid-infrared (MIR). The waveguide consists of a rectangular core supporting a low-loss optical mode, suspended in air by a series of transverse ribs. The ribs are patterned to form a finite quasi-one-dimensional phononic crystal, with the complete stopband suppressing the transverse leakage of acoustic waves, and confining them to the core of the waveguide. We derive a theoretical formalism that can be used to compute the opto-acoustic interaction in such periodic structures, and find forward intramodal-SBS gains up to $1750~\text{m}^{-1}\text{W}^{-1}$, which compares favorably with the proposed MIR SBS designs based on buried germanium waveguides. This large gain is achieved thanks to the nearly complete suppression of acoustic radiative losses.
	\end{abstract}		
	\maketitle

\section{Introduction}
Stimulated Brillouin Scattering (SBS), which describes the coherent nonlinear interaction between optical and acoustic fields\cite{boydbook, eggleton2013inducing}, is a key effect for a wide range of photonics capabilities, including wideband tunable, ultra-narrow RF filters \cite{zhang2011widely, marpaung2015low}, acousto-optical storage\cite{zhu2007stored,birgitrecent}, non-reciprocal photonic elements\cite{sounas2017non} and new laser sources\cite{otterstrom2018silicon}. The ability to generate a useful level of SBS gain in a short waveguide is especially important in the mid-IR, where there is particular demand for broadband, tuneable filters for spectroscopy or IR sensors \cite{Wolff:14,soref2010mid}. Furthermore, by migrating nonlinear photonics towards mid-infrared range, the unwanted two-photon absorption (TPA) in the two key CMOS compatible materials: silicon and germanium, can be eliminated\cite{soref2010mid,doi:10.1063/1.3592270}.

A central challenge in harnessing SBS is to design a waveguide which confines both the optical and acoustic waves. The obvious approach to this task is to confine both waves using total internal reflection (TIR) --- this approach requires materials with both a high refractive index and low stiffness, and while realizations of this scheme have been reported, they are limited to a small range of materials\cite{Pant:11} that require specialized fabrication techniques. TIR can also be achieved by \textit{geometric softening} of the guided acoustic modes\cite{doi:10.1063/1.4955002} to reduce their phase velocities below that of the substrate and surface waves, thereby prohibiting acoustic loss. Another class of strategies relies on geometric isolation of the acoustic modes from the substrate, for example,
by designing suspended waveguides with few, spatially-separated supports \cite{shin2013tailorable,kittlaus2016large,van2015net}, or using phoxonic crystals \cite{doi:10.1063/1.2216885,zhang2017design} which guide both photons and phonons along line defects. Each of these strategies has advantages and drawbacks. Two-dimensional phoxonic crystals offer a unique control over the propagation and co-localization of photons and phonons along line and point defects, but they require simultaneous designing of optical and acoustic bandgaps. Suspended structures, while simpler to design and fabricate, inevitably suffer from losses through the points of contact, which are rigidly clamped to the substrate and allow radiative loss of the acoustic mode \cite{van2015net}. 

Here we propose a new class of silicon suspended structures which can be used to achieve high forward SBS gains of up to 1750~(mW)$^{-1}$ over broad bandwidths in 
the mid-IR. The structure achieves acoustic isolation via a combination of TIR and
geometrical shielding of the acoustic modes: the central idea is to confine the acoustic modes by suspending the waveguide with an array of flexible ribs, with each rib structured to induce a phononic stopband in the transverse direction. By tuning the stopband to the frequency of the acoustic mode we suppress the transmission of acoustic waves into the substrate through the ribs, and simultaneously reduce the clamping losses\cite{yu2012control}. This strategy, implemented recently in two-dimensional membranes\cite{tsaturyan2017ultracoherent} and nanobeams\cite{Ghadimi764} provides silicon acoustic waveguides with mechanical quality factors $Q_m$ close to 90\% of those found for unsupported acoustic waveguides. Simultaneously, the ribs form a subwavelength grating (SWG) for mid-IR light propagating in the waveguide. Such a grating can be seen as an effective cladding layer with refractive index very close to unity, and can provide TIR guidance for the optical field. 

We present here a proof-of-concept design for this new structure, and provide design rules for the extension 
of the concept to different regimes of wavelength and material parameters. We discuss the mechanisms of optical guidance in these waveguides and formulate constraints on the structure's geometric characteristics, such as the required spacing between suspending ribs. We discuss the creation of acoustic stopband at the frequency of the mechanical vibrations of the waveguide by patterning of the ribs, compute the resulting acoustic confinement and investigate the geometrical dependence of
the acoustic loss. This results in a set of design guidelines for creating these types of suspended, softly-clamped waveguides for SBS applications over a broad range of wavelengths. Finally, we present the formalism 
for Brillouin gain computations in a periodic system, and use this to estimate SBS gain 
for a realistic silicon platform. We find that these structures exhibit gains that are comparable with the predicted gains for 
mid-IR structures in germanium\cite{Wolff:14}, and so represent a viable alternative for harnessing SBS in this spectral range.

\subsection{Suspended MIR waveguides}

\begin{figure*}[htbp!]
	\begin{center}
		\includegraphics[width=.75\textwidth]{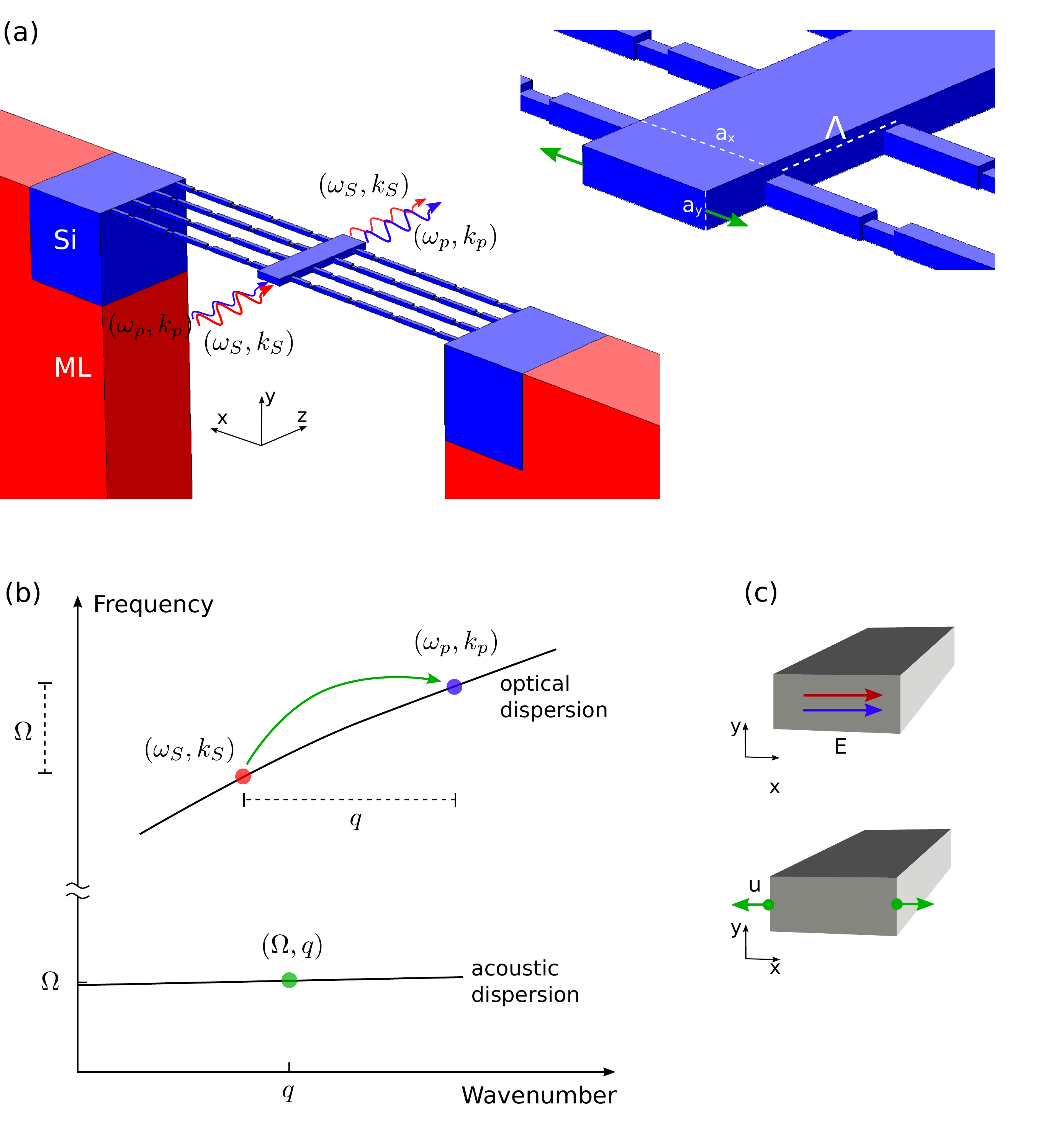}
		\caption{(a) A section of the suspended waveguide setup, with 4 unit cells along $z$ axis, and a zoomed in section of the waveguide. Patterned ribs, spaced by $\Lambda$, connect the central waveguide with rectangular cross section $(a_x\times a_y)$, with the slab region. While the entire structure is made of silicon (Si), Matching Layers (ML) are introduced in the slab region, using built-in COMSOL procedures for optics \cite{secondcomsol}, and following the method described in Ref.~\citenum{steeneken2013parameter} for acoustics. Unless otherwise specified, throughout this work we will be discussing structures with $(a_x, a_y, \Lambda)=(1.2,0.3, 1.25)~\mu$m (see the discussion in the main text), and ribs with square cross section $(0.18~\mu\text{m})^2$. The patterning of ribs is discussed in Section \ref{acoustic.response}. (b) Phase-matching in Forward intramodal Stimulated Brillouin Scattering (intra-mode FBS). Coupling between the optical pump and Stokes waves is mediated by the co-propagating acoustic wave, fulfilling the phase-matching conditions $k_S+q=k_p$ and $\omega_S+\Omega=\omega_p$. (c) Schematic representations of the fundamental optical TE fields of the pump and Stokes waves (upper panel), and the lowest-order symmetry-allowed acoustic mode \cite{Wolffsymmetries}, associated with lateral stretching of the waveguide (lower panel).}
		\label{Fig_schematic}
	\end{center}
\end{figure*}

\begin{figure*}[htbp!]
	\begin{center}
		\includegraphics[width=.75\textwidth]{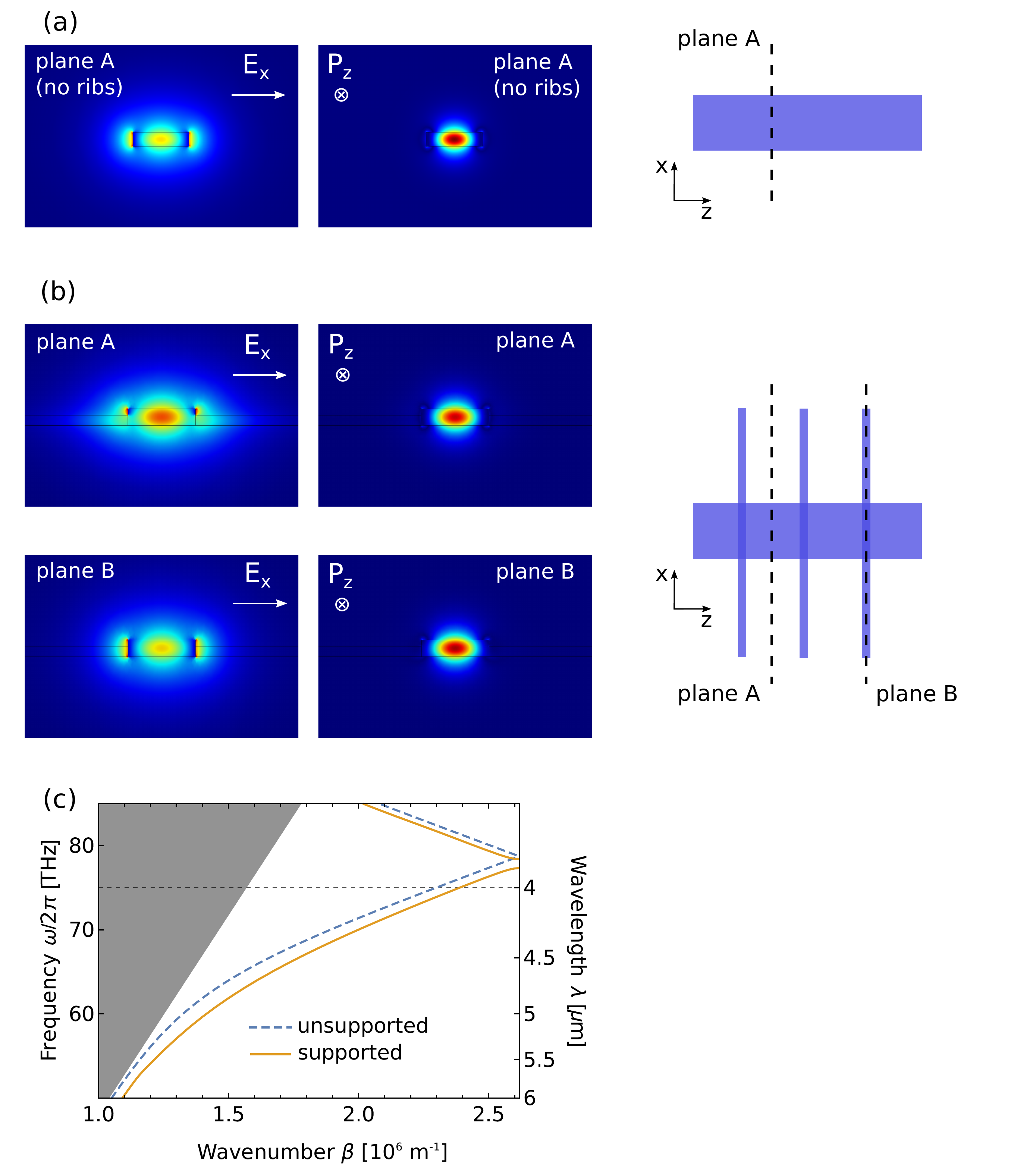}
		\caption{Optical waveguiding inside the central bar. (a,b) Spatial distribution of the dominant $E_x$ component of the electric field (upper panel) and the $P_z$ component of the Poynting vector (lower panel) in the planes marked in the panels on the right. (c) Dispersion relation of the rectangular silicon waveguide with dimensions $(1.2, 0.3)~\mu$m, unsupported (dashed blue line) and supported by the unpatterned ribs spaced by $\Lambda=1.2~\mu$m (solid orange line). The horizontal line denotes the $\lambda_0=4~\mu$m waveguiding wavelength.}
		\label{Fig_optics}
	\end{center}
\end{figure*}

Suspended waveguides for low-loss mid-IR light guiding in silicon rely on the suspending ribs forming a subwavelength grating (SWG).\cite{SolerPenades:14,penades2016suspended} %We show schematically an example of such a waveguide in Fig. 1(a). 
In SWG guidance, the spacing between ribs (the pitch of the structure $\Lambda$, see schematics in Fig.~\ref{Fig_schematic}(a)) must be smaller than half of the effective wavelength of light $\lambda_0/\neff$. %This condition formulates a tradeoff between the optical confinement of the mode (expressed through \neff), and vibrational isolation of the waveguide, governed by $\Lambda$. 
While in general the optical dispersion relation of the waveguide depends on $\Lambda$, here we assume that the ribs do not strongly modify the optical response of the structure, and estimate the upper limit for $\Lambda$ from the dispersion relation of the unsupported waveguide. For example, for a rectangular silicon waveguide in air similar to the one reported by Penandes \textit{et al.}\cite{penades2016suspended}, and shown in Fig.~\ref{Fig_optics}(a), with cross section dimensions $(1.2, 0.3)~\mu$m and operating at $4~\mu$m mid-IR wavelength, we find for the fundamental TE mode $\neff \approx 1.4$ and, consequently, $\Lambda < 1.3~\mu$m. Therefore, throughout this work, we will consider suspended waveguides with pitch  $\Lambda=1.25~\mu$m. We should note that compared to the structure discussed by Penandes \textit{et al.}\cite{penades2016suspended}, our waveguide has a significantly larger aspect ratio, exhibits a significantly lower effective index $\neff$ of the fundamental mode, and consequently, puts less stringent condition on the pitch $\Lambda$.

Comparison of calculations for a waveguiding structure with and without the ribs (see Fig.~\ref{Fig_optics}) confirms that the transverse supporting structures introduce a very small modification to the optical response of the waveguide. In particular, the calculated dispersion relation (orange lines in Fig.~\ref{Fig_optics}(c)) and field profiles (Fig.~\ref{Fig_optics}(b)) of the TE mode in the suspended structure with simplified, unstructured square ribs with (0.18~$\mu\text{m})^2$ cross section (see schematic in Fig.~\ref{Fig_optics}(a)) follow closely those found for unsuspended structures (see Fig.~\ref{Fig_optics}(a)). The particular design of the ribs (i.e. patterning) does not influence the guiding properties significantly, because light cannot be efficiently guided out of the central bar through the ribs. On the other hand, the ribs do induce a small change of the average refractive index of the environment, red-shifting the dispersion relation slightly (Fig.~\ref{Fig_optics}(c)). Furthermore, they form a strong Bragg grating, and introduce a partial photonic stopband of width $1$~THz, centered at $77.9$~THz - an effect which could be used to further enhance, or suppress the Brillouin gain \cite{merklein2015enhancing}.

We have characterized the optical response of unsupported and supported waveguides by implementing 2D and 3D models, respectively, in the RF module of the COMSOL software \cite{secondcomsol}. The refractive index of silicon in the mid-IR was taken as constant $n=3.42$,\cite{li1980refractive} and perfectly matching layers were used in the slab region (marked in red in Fig.~\ref{Fig_schematic}(a)). In 3D systems, we applied Floquet boundary conditions along the $\hat{z}$ axis with period $\Lambda$.

\section{Acoustic response}\label{acoustic.response}

\begin{figure*}[htbp!]
	\begin{center}
		\includegraphics[width=.75\textwidth]{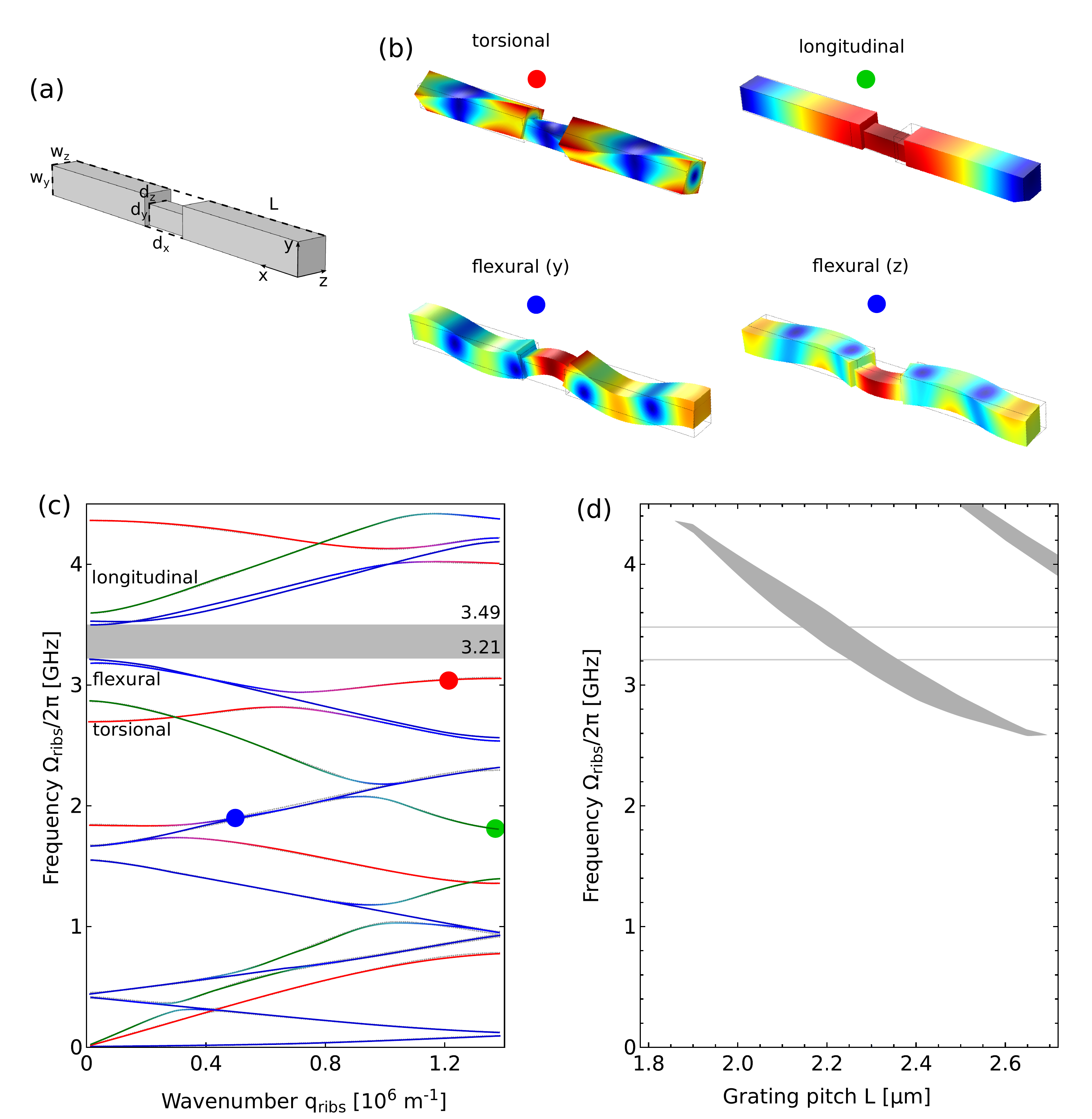}
		\caption{Acoustic response of the phononic crystal forming the patterned ribs. (a) Schematic of a unit cell of the phononic crystal. (b) Acoustic modes of the one-dimensional phononic crystal, calculated at positions marked with colored dots in (c). (c) Dispersion diagram of the one-dimensional phononic crystals for periodicity $L=2.25~\mu$m and $(d_x,d_y,d_z)=(340,120,120)$~nm, patterned in the rib with square cross-section $(w_y,w_z)=(180,180)$~nm, used throughout the rest of the work. Green, blue and red lines denote the longitudinal (compressional), flexural and torsional modes, marked on the dispersion diagrams with circles. Hybridization is marked by intermediate colors. Lines are guides only. The complete phononic stopband is centered at 3.35~GHz, and can be tuned by changing the length of the unit cell $L$, as shown in (d), where we mark the stopbands as shaded areas.}
		\label{Fig_acoustics}
	\end{center}
\end{figure*}

Acoustic response of the suspended structure can be largely controlled by engineering the capabilities of ribs to guide elastic waves away from the central waveguide. In particular, by patterning the ribs, we can form a complete acoustic stopband, and forbid acoustic waves from dissipating through the ribs. We illustrate this concept in Fig.~\ref{Fig_acoustics} by considering an infinite one-dimensional phononic crystal forming the ribs, periodic along the $\hat{x}$ axis, with the unit cell shown schematically in Fig.~\ref{Fig_acoustics}(a). The asymmetry along $\hat{x}$ in the grating structure is included to simplify the fabrication process, and induces the splitting of flexural modes polarized along $\hat{y}$ and $\hat{z}$ directions (see schematic in Fig.~\ref{Fig_acoustics}(b)). In Fig.~\ref{Fig_acoustics}(c), we plot the dispersion diagram ($q_{\text{ribs}}$, $\Omega_{\text{ribs}}$) of the particular design of patterned ribs ($(d_x,d_y,d_z,L)=(340,120,120,2250)$~nm), which we use throughout the rest of this work. This plot reveals a complete stopband centered at $3.35$~GHz, with the $0.28$~GHz width determined by the flexural modes (blue lines) shown in the bottom row of Fig.~\ref{Fig_acoustics}(b). The complete stopband of the one-dimensional phononic crystal can be tuned over a broad spectral range, e.g.\ by changing the length $L$ of the unit cell, as shown in Fig.~\ref{Fig_acoustics}(d).

All the numerical calculations of mechanical response were carried out using the Structural Mechanics module of the COMSOL software \cite{secondcomsol}. Silicon was described by stiffness and acoustic loss cubic tensors with numerical values taken from Ref. \citenum{Smith:16}, with the principial axes of the crystal coinciding with the $\hat{x}\hat{y}\hat{z}$ axes. In the calculations of the response of the entire structure, we include elastic matching layers, marked in Fig.~\ref{Fig_schematic}(a) as red volumes, implementing the method described in Ref. \citenum{steeneken2013parameter}. The periodicity --- both of the phononic crystal forming ribs (along axis $\hat{x}$), as well as the entire waveguiding system (along axis $\hat{z}$) --- was accounted for by imposing Floquet boundary conditions along the direction of the periodicity.

\begin{figure*}[htbp!]
	\begin{center}
		\includegraphics[width=.75\textwidth]{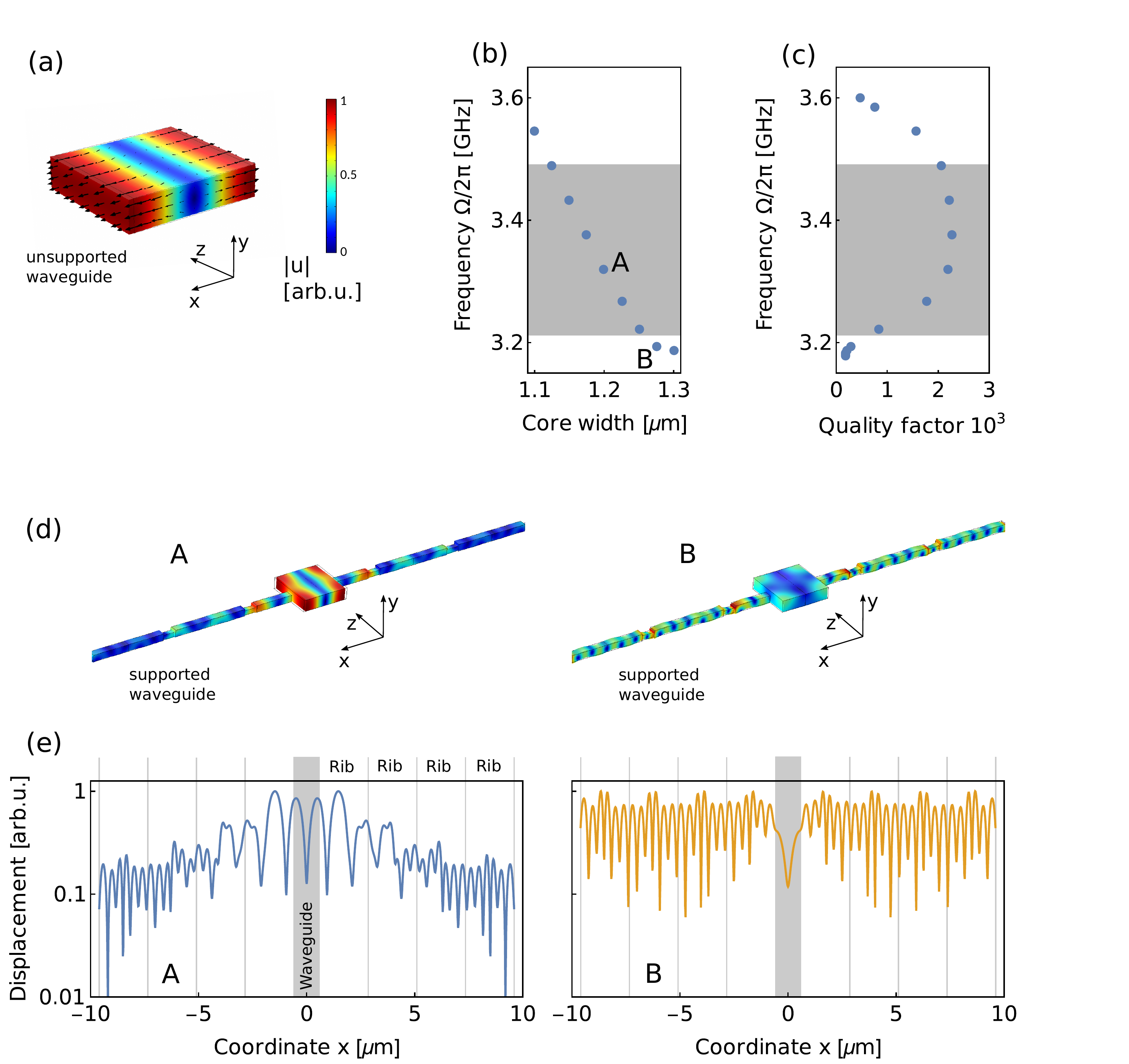}
		\caption{(a) Displacement field distribution of a lateral stretching mode of an unsupported waveguide. (b) Frequency and (c) mechanical quality factor $Q_m$ of the waveguide suspended by patterned ribs as a function of the core width $a_x$. Stopband of the phononic crystal is marked as the gray area. The ribs comprise 4 unit cells of the phononic crystal on each side of the waveguide. (d,e) Schematics of the displacement fields in the unit cell, associated with modes marked in (a) as A and B. Plots in (e) demonstrate the displacement field calculated along the $x$ axis, passing through the centres of the ribs. For the mode A with resonance within the stopband the displacement field decays exponentially over the length of the unit cells of the phononic crystal.}
		\label{Fig_mechanical}
	\end{center}
\end{figure*}

We can now consider the elastic mode of the entire waveguiding structure, which mediates the SBS interaction between two optical waves in TE modes propagating in the suspended waveguide. We choose to study forward intramodal SBS (intra-mode FBS, see Fig. \ref{Fig_schematic}(b)), in which mechanical modes characterized by longitudinal wavenumber and frequency $(q,\Omega)$ mediate the interaction between co-propagating optical beams ({pump} and {Stokes}, denoted by subscripts $p$ and $S$, respectively) characterized by $(k_p,\omega_p)$ and $(k_S,\omega_S)$. From the phase matching conditions, we find that the magnitude of the mechanical wavenumber $q$ is given approximately by $q = k_S-k_p \approx (\omega_S-\omega_p)\neff/c = \Omega\neff/c$, where $\neff$ is the optical mode index near $\omega_p$. Since the typical vibrational frequencies are of the order of GHz, we can take $q=0$. Furthermore, we focus on the lowest-order acoustic mode associated with lateral stretching mode of the waveguide, depicted schematically in Fig.~\ref{Fig_schematic}(c)\cite{shin2013tailorable,van2015net}. For the simplest, though experimentally unfeasible, unsupported waveguide, we have found the frequency ($\Omega = 3.69$~GHz), mechanical quality factor ($Q_b = 2970$, limited by the viscosity of silicon) and displacement field (Fig.~\ref{Fig_mechanical}(a)) of that mode.

Apart from the viscous losses in silicon, the mechanical quality factor of a more realistic, supported structure is, as discussed earlier, determined by the ability of the ribs to guide acoustic waves into the substrate (or \textit{slab region}). We can demonstrate that effect by detuning the acoustic mode of the waveguide across the stopband of the phononic crystal, and calculating the mechanical quality factor $Q_b$ of the entire structure (accounting for both the viscosity and dissipation into the slab region). To this end, we change the width $a_x$ of the central core (see Fig.~\ref{Fig_schematic}(a))), and find the frequency $\Omega$ (Fig.~\ref{Fig_mechanical}(b)) and quality factor $Q_b$ (Fig.~\ref{Fig_mechanical}(c)) of the mode. These calculations are carried out assuming that the ribs include 4 unit cells of the one-dimensional phononic crystal discussed earlier. As the core width increases, the mode frequency redshifts and passes through the stopband of the phononic crystal. For a system with resonance inside the stopband (i.e. between $3.21$ and $3.48$~GHz), the displacement field is localized inside the central waveguide, and the energy does not propagate towards the clamps (see the displacement field distribution at point A, calculated for $a_x=1.2~\mu$m, shown in Fig.~\ref{Fig_mechanical}(d,e)). Consequently, the system exhibits large quality factors, comparable to those of the unsupported waveguide. As we increase the core width, the resonances shift outside of the stopband, and the quality factor drops rapidly. For this structure, the ribs oscillate along their entire length and, through clamping, transfer the energy into the slab region. An example displacement field distribution, calculated for $a_x=1.275~\mu$m, is shown in panel B in Fig.~\ref{Fig_mechanical}(d,e). We should note that the effect of suppression of energy dissipation is not limited to the particular lateral stretching mode discussed above, but should be observed for any acoustic mode of the central waveguide, tuned to the stopband.

Alternatively, the suppression of acoustic dissipation could be attributed to the reduced transmission of mechanical energy through the narrower parts of the structured ribs. We can dismiss this explanation by noting that the three families of elastic modes carrying energy through the ribs exhibit no cutoff, and would be supported also by the thin sections of the ribs.

\section{Estimating FBS gain}\label{calculations}

We now calculate the Brillouin gain $\Gamma$ in the suspended structure, by extending the formalism originally developed to treat translationally invariant systems, introduced by Wolff \textit{et al.}\cite{wolff2015stimulated}. To this end we consider the Bloch picture of the acoustic modes of the quasi-one-dimensional system:
\begin{equation}\label{Ubloch}
\mathbf{U}(\mathbf{r}) = b(z) {\mathbf{u}}_q (\mathbf{r})= b(z) \tilde{\mathbf{u}}_q (\mathbf{r}) e^{iq z},
\end{equation}
where $\tilde{\mathbf{u}}$ is a periodic function along the $z$ coordinate
\begin{equation}
\tilde{\mathbf{u}}_q(\mathbf{r}+L \hat{z}) = \tilde{\mathbf{u}}_q (\mathbf{r}),
\end{equation} 
and $b(z)$ is a slowly varying envelope with $\partial_z b\ll L^{-1}$. Similarly, we write down the electric fields for the two optical modes (pump $\mathbf{E}^{(1)}$ and Stokes $\mathbf{E}^{(2)}$) as
\begin{equation}\label{Ebloch}
\mathbf{E}^{(i)}(\mathbf{r}) = a^{(i)}(z)\mathbf{e}^{(i)}_{k^{(i)}} (\mathbf{r})= a^{(i)}(z)\tilde{\mathbf{e}}^{(i)}_{k^{(i)}} (\mathbf{r}) e^{i {k^{(i)}} z},
\end{equation}
with $\mathbf{e}^{(i)}_{k^{(i)}}(\mathbf{r}+a \hat{z}) = \mathbf{e}^{(i)}_{k^{(i)}} (\mathbf{r})$, and $\partial_z a\ll L^{-1}$. In intra-mode FBS, we consider the pump and Stokes beams to propagate in the same mode. In Appendix \ref{AppendixA} we provide a full derivation of the approximate Brillouin under perfect phase-matching $k^{(2)}+q=k^{(1)}$ condition:
\begin{equation}\label{Bgain2_maintext}
\Gamma \approx 4\omega \frac{Q_m \left|\left\langle \QPEa+\QMB\right\rangle\right|^2}{\langle{\mathcal{E}_b} \rangle \langle \mathcal{P}^{(1)}\rangle \langle \mathcal{P}^{(2)}\rangle},
\end{equation}
where the averages $\langle . \rangle$ are carried out over the unit cell, e.g. $\langle \mathcal{P}^{(i)}\rangle$ describes the $z$ component of the average flux of optical energy through the waveguide (see Eq.~(\ref{Pi},\ref{dynamics.a01})) and $\langle{\mathcal{E}_b} \rangle$ describes the acoustic energy density (Eq.~(\ref{Eb})). The overlap integral between unnormalized optical and acoustic modes $\QPEa$ due to the photoelastic effect is defined by
\begin{equation}
\QPEa(z) = -\varepsilon_0 \int \text{d}^2r \varepsilon_a^2 \sum_{ijkl}  [e_i^{(1)}]^*e_j^{(2)} p_{ijkl} \partial_k u_l^*,
\end{equation}
where $p_{ijkl}$ is the Pockels tensor, and integration is carried out in a $\hat{x}\hat{y}$ plane, determined by argument $z$. The effect of moving boundaries is expressed through overlap integral $\QMB(z)$ calculated as integral over the boundaries between materials (with relative permittivities $\varepsilon_a$ and $\varepsilon_b$) in the same $\hat{x}\hat{y}$ plane:
\begin{align}
\QMB(z) = \int_A \text{d}\mathbf{r} &(\mathbf{u}^* \cdot \hat{n})\left[(\varepsilon_a-\varepsilon_b)\varepsilon_0\left(\hat{n}\times \mathbf{e}^{(1)}\right)^*\left(\hat{n}\times \mathbf{e}^{(2)}\right) - \left(\frac{1}{\varepsilon_a}-\frac{1}{\varepsilon_b}\right)\frac{1}{\varepsilon_0}\left(\hat{n}\cdot\mathbf{d}^{(1)}\right)^*\left(\hat{n}\cdot\mathbf{d}^{(2)}\right)\right].
\end{align}

\begin{figure*}[htbp!]
	\begin{center}
		\includegraphics[width=.75\textwidth]{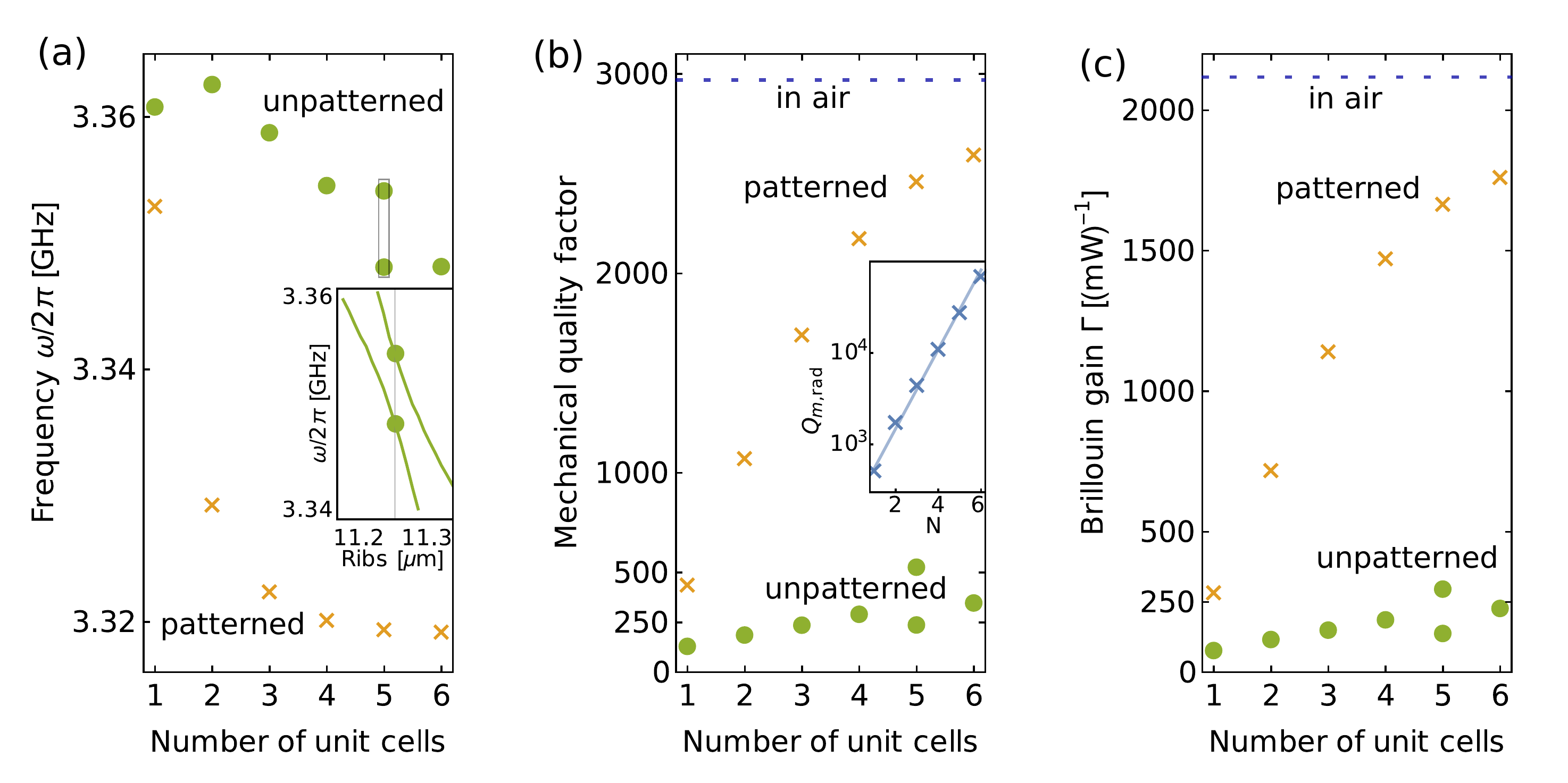}
		\caption{Acoustic frequencies (a), mechanical quality factors $Q_m$ (b) and Brillouin gain coefficients $\Gamma$ (c) of suspended waveguide systems, calculated for the ribs comprised of an increasing number of unit cells of the phononic crystal. Orange crosses and green dots represent systems with patterned, and unpatterned ribs, respectively. Inset in (a) traces two anti-crossing mechanical modes, while the inset in (b) shows the calculated radiative mechanical quality factor $Q_{m,\text{rad}}$ of patterned waveguide, as a function of the ribs length, with an exponential fit. The dashed blue lines in (b) and (c) denote the values obtained for unsupported waveguides.}
		\label{Fig_FSBS}
	\end{center}
\end{figure*}

Results of the calculation of Brillouin gain are shown in Fig.~\ref{Fig_FSBS}, as a function of the number $N$ of unit cells (alternatively, length of the ribs $N\times\Lambda$). As the ribs become longer, the mechanical frequencies (a) of structures with patterned (orange crosses) and unpatterned ribs (green dots) decrease. As we show in the inset, the peculiar splitting of modes in the latter case is due to the anti-crossing between two modes of structure with unpatterned ribs of length around $5\Lambda = 11.25~\mu$m. Simultaneously, the mechanical quality factors $Q_m$ (b) grow from 440 (140) to 2600 (360) for patterned (unpatterned) structures. This change, by factors of about 6 (2.5) is mostly responsible for the simultaneous increase in the Brillouin gain coefficient $\Gamma$ shown in Fig. \ref{Fig_FSBS}(c). This confirms our expectation that, since the optical field is largely confined  inside and near the waveguide, neither the overlap integrals $\QPEa$ and $\QMB$, nor the averaged optical fluxes $\langle \mathcal{P}^{(i)}\rangle$ depend strongly on the length of the ribs. Furthermore, the average acoustic energy density $\langle{\mathcal{E}_b} \rangle$ depends very weakly on the rib length due to their small volume - even in the case of unpatterned systems. We also find that the two contributions to the Brillouin gain - photoelasticity and radiation pressure - retain the similar ratio for every investigated structure $\langle\QPEa\rangle / \langle\QMB\rangle \approx 0.6$. These observations simplify Eq.~(\ref{Bgain2_maintext}) to $\Gamma \propto Q_m$, a relationship we recover in Fig.~\ref{Fig_FSBS}.

Furthermore, by tracing the dependence of $Q_m$ on the number of patterned ribs, we can estimate the non-radiative contribution to the acoustic decay. If we insist that the radiative mechanical quality factor $Q_{m,\text{rad}}$ should grow exponentially with the number of ribs\cite{kalaee2018quantum} $N$, we can estimate the mechanical quality factor due to the viscous losses $Q_{m,\text{visc}}$ from
\begin{equation}
	Q_m^{-1}=Q_{m,\text{visc}}^{-1}+Q_{m,\text{rad}}^{-1}.
\end{equation}
Fitting the dependence of $Q_{m,\text{rad}}$ on $N$ to the numerical results (see inset in Fig.~\ref{Fig_FSBS}(b)), we find $Q_{m,\text{visc}}\approx 2700$. The difference between this magnitude and the quality factor $Q_{m,0}=2970$ of the unsupported structure (marked as blue dashed line in Fig.~\ref{Fig_FSBS}(b)) quantifies the viscous losses in the ribs.

The mechanical quality factors and Brillouin gain coefficients we discuss above compare favorably with those reported for the few realistic designs for near- and mid-IR systems proposed to date. These include germanium waveguides buried in silicon nitrade \cite{Wolff:14, DeLeonardis:16} operating at $4~\mu$m, which enable backward-SBS with similar mechanical quality factors, and Brillouin gain up to $1000~\text{m}^{-1}\text{W}^{-1}$. In an experimental realization of a rib waveguide\cite{Pant:11} operating in near-IR, backward-SBS gain coefficients was reported as about $300~\text{m}^{-1}\text{W}^{-1}$.

\section{Summary and outlook}

We have proposed a novel type of silicon waveguides capable of supporting both low-loss MIR optical and GHz acoustic waves. Our design is based on previous proposals for optical subwavelength guidance in silicon waveguides suspended in air by periodic ribs. To simultaneously confine the acoustic waves inside the waveguide, we structured the supporting ribs to exhibit a complete acoustic stopband. The mechanical quality factor of such structures can reach about 90\% of the viscosity-limited quality factor of an unsupported waveguide, indicating that we can almost completely eliminate the dissipation of acoustic waves into the slab region. This isolation also boosts the forward intramodal Brillouin gain coefficient, which can reach $1750~\text{m}^{-1}\text{W}^{-1}$.

This design can be further refined to explore its applicability to the backwards SBS, or the efficiency of acoustic isolation through 1D phononic crystals with partial stopband (see e.g. \cite{Ghadimi764}). Besides further enhancing the Brillouin gain, enhanced control over the channels of acoustic dissipation and propagation might also pave the way to designing novel acoustic beam splitters or couplers.

\newpage
\appendix

\section{Forward Stimulated Brillouin Scattering in periodic structures}\label{AppendixA}
In a periodic opto-acoustic system, such as the suspended waveguide discussed in this work, Brillouin gain can be calculated by adopting a Bloch picture mode for the quasi-1D system with period $L$. In this formalism, we write the displacement field and the electric field in the Bloch form given in Eqs.~(\ref{Ubloch}, \ref{Ebloch}).
For simplicity, we drop subscripts $q$ and $\beta$ characterizing wave numbers. Furthermore, while in this contribution we focus on FSBS (where $q\approx 0$), the derivation shown below will be general, as to be applicable to BSBS (where $q\approx 2\beta$).

Modes $\mathbf{u}_q (\mathbf{r})$ and $\mathbf{e}^{(i)}_\beta (\mathbf{r})$ can be found by solving linear elastic, and Maxwell equations, respectively, by enforcing Floquet boundary conditions to the $z$-normal faces of a unit cell.

\subsection{Formulating dynamical equations}
 
To derive the expression for the Brillouin gain in this periodic structure, let us revisit the corresponding derivation for a waveguide-like, translationally-invariant system where functions $\tilde{\mathbf{u}}(\mathbf{r})$ and $\tilde{\mathbf{e}}(\mathbf{r})$ are functions of transverse $\mathbf{r}_\perp = (x,y)$ coordinates only.

For the periodic structure, the electromagnetic energy density and the $z$ component of the energy flux for a fixed $z$ coordinate can be calculated as:
\begin{equation}
	\mathcal{E}^{(i)}(z) = 2 \varepsilon_0 \int \text{d}^2r \varepsilon(\mathbf{r})[\mathbf{e}^{(i)}(\mathbf{r})]^* \cdot \mathbf{e}^{(i)}(\mathbf{r}),
\end{equation}
\begin{equation}\label{Pi}
	\mathcal{P}^{(i)}(z) = 2 \int \text{d}^2r \hat{z} \cdot ([\mathbf{e}^{(i)}(\mathbf{r})]^* \times \mathbf{h}^{(i)}(\mathbf{r})).
\end{equation}
where the integration is carried in the transverse $\hat{x}\hat{y}$ plane. It should be noted that in the discussed system, the flux density will have small, but non-vanishing components in $\hat{x}\hat{y}$ plane associated with the energy leaking out from the waveguide into radiation modes. Nevertheless, for systems optimized to serve as low-loss optical waveguides, these terms should be negligible.

In the limit of longitudinally-invariant structures, $\mathcal{E}^{(i)}$ and $\mathcal{P}^{(i)}$ are constant, and their ratio describes the \textit{energy transport velocity of the mode} (energy velocity) \cite{wolff2015stimulated} \begin{equation}\label{vE1}
	v_E = \frac{\mathcal{P}^{(i)}}{\mathcal{E}^{(i)}}.
\end{equation}
In the absence of material losses, this is equal to the group velocity.

For a periodic structure, the energy velocity $v_E$ can be obtained by separately averaging these magnitudes over the volume of the unit cell (UC) \cite{chen2010group}
\begin{equation}\label{vE2}
	v_E = \frac{\langle \mathcal{P}^{(i)}(z) \rangle}{\langle \mathcal{E}^{(i)}(z) \rangle}.
\end{equation}
In a lossless system, this velocity is again equal to the group velocity of the mode.

We have verified that both Eqs.~(\ref{vE1}, \ref{vE2}) provide good estimates of the group velocity read out from the dispersion relations of the waveguides without and with ribs, respectively, discussed in Fig. \ref{Fig_optics}.

\subsubsection{Effective dynamics of optical envelopes}

Using these definitions of energy density and flux, we can repeat the entire derivation presented by Wolff \textit{et al.}\cite{wolff2015stimulated} up to Eq.~(26):
\begin{align}\label{dynamics.a0}
-i\omega^{(1)} \left[\mathcal{P}^{(1)}(z) \partial_z a^{(1)}(z) + \mathcal{E}^{(1)}(z) \partial_t a^{(1)}(z) \right] = a^{(2)}(z) b^*(z)\left(-i\omega^{(1)}\right)^2\mathcal{Q}_1(z),
\end{align}
where
\begin{equation}
	\mathcal{Q}_i(z) = \int \text{d}^2r \left[[\mathbf{e}^{(i)}(\mathbf{r})]^* \cdot \Delta \mathbf{d}^{(i)}(\mathbf{r}) - [\mathbf{d}^{(i)}(\mathbf{r})]^* \cdot \Delta \mathbf{e}^{(i)}(\mathbf{r})-\mu_0 [\mathbf{h}^{(i)}(\mathbf{r})]^* \cdot \Delta \mathbf{h}^{(i)}(\mathbf{r})\right].
\end{equation} 
Here, fields denoted with $\Delta$ describe the thus-far-unspecified perturbations giving rise to the coupling between the optical and acoustic fields. The photoelastic and moving boundary effects governing this coupling are discussed in section \ref{calculations} in the main text.
%Throughout this manuscript, we focus on the contributions to coupling from the photoelastic effect
%\begin{equation}
%\QPEa(z) = -\varepsilon_0 \int \text{d}^2r [\varepsilon(\mathbf{r})]^2 \sum_{ijkl}  [e_i^{(1)}(\mathbf{r})]^*e_j^{(2)}(\mathbf{r}) p_{ijkl} \partial_k u_l^*(\mathbf{r}),
%\end{equation}
%where the integration is carried out over the $xy$ plane for a fixed coordinate $z$.

Equation (\ref{dynamics.a0}) mixes the slow evolution of the envelope functions $a^{(i)}(z)$ and $b(z)$ with the rapidly changing, periodic functions $\mathcal{P}^{(i)}(z)$, $\mathcal{E}^{(i)}(z)$ and $\mathcal{Q}_1(z)$ defined by the Bloch modes of the system. We can separate the two, to arrive at the evolution equations for the envelopes, by averaging both sides of Eq.~(\ref{dynamics.a0}) over the length of the unit cell $L$, and assuming that over that distance the envelopes are almost constant
\begin{align}\label{dynamics.a01}
\int_z^{z+L} \text{d}z' \left[\partial_z a^{(1)}(z)\right]_{z=z'} \mathcal{P}^{(1)}(z')  \approx \partial_z a^{(1)}(z) \int_z^{z+L} \text{d}z' \mathcal{P}^{(1)}(z') = \partial_z a^{(1)}(z) \langle \mathcal{P}^{(1)}\rangle L,
\end{align}
arriving at:
\begin{align}\label{dynamics.a1}
	\partial_z a^{(1)}(z) + \frac{1}{v^{(1)}} \partial_t a^{(1)}(z) = -i\omega^{(1)}a^{(2)}(z) b^*(z)\frac{\langle \mathcal{Q}_1\rangle}{\langle \mathcal{P}^{(1)}\rangle},
\end{align}
where $v^{(1)}$ is the energy velocity of the mode defined in Eq.~(\ref{vE2}).

The analogous equation for the other optical envelope reads
\begin{align}\label{dynamics.a11}
\partial_z a^{(2)}(z) + \frac{1}{v^{(2)}} \partial_t a^{(2)}(z) = -i\omega^{(2)}a^{(1)}(z) b(z)\frac{\langle \mathcal{Q}_2\rangle}{\langle \mathcal{P}^{(2)}\rangle},
\end{align}
where the velocity $v^{(2)}$ and integrated flux $\langle \mathcal{P}^{(2)}\rangle$ are defined similarly as for the first optical mode. From the definition of the PE contribution to the overlap integral $\QPEa(z)$, one can directly find that 
$\QPEa(z)= \left[\QPEb(z)\right]^*$.

\subsubsection{Effective dynamics of acoustic envelopes}

To derive the dynamic equations for the vibrations, we take Eq.~(43) from \cite{wolff2015stimulated}:
\begin{equation}\label{dynamics.b1}
	-i\Omega \sum_{jkl} [(c_{izkl}\partial_k+\partial_j c_{ijzl})u_l(\mathbf{r}) \partial_z b(z) -2i \Omega \rho(\mathbf{r}) u_i(\mathbf{r}) \partial_t b(z)+ (\partial_j \eta_{ijkl}\partial_k u_l(\mathbf{r}))b(z) +  [a^{(1)}(z)]^* a^{(2)}(z) f_i]+c.c.=0.
\end{equation}
Multiplying both sides by $\mathbf{u}^*$, integrating over the transverse plane
and using the definitions of acoustic energy density and flux:
\begin{equation}\label{Eb}
 \mathcal{E}_b(z) = 2\Omega^2 \int \text{d}^2r \rho(\mathbf{r})|\mathbf{u}(\mathbf{r})|^2,\quad  \mathcal{P}_b(z) = -2i\Omega \int \text{d}^2r \sum_{ikl}c_{zikl}[u_i(\mathbf{r})]^* \partial_k u_l(\mathbf{r}),
\end{equation}
we arrive at
\begin{align}\label{dynamics.b2}
\mathcal{P}_{b}(z) \partial_z b(z) + \mathcal{E}_{b}(z) \partial_t b(z) + b(z)\alpha_{\mathcal{P}_b}(z) = -i \Omega [a^{(1)}(z)]^* a^{(2)}(z)\mathcal{Q}_b(z),
\end{align}
where
\begin{equation}
	\mathcal{Q}_b(z) = \int \text{d}^2r [\mathbf{u}(x,y;z)]^* \cdot \mathbf{f}(x,y;z)
\end{equation}
governs the coupling between the acoustic field and applied force density $\mathbf{f}$, and 
\begin{equation}\label{alphaP}
	\alpha_{\mathcal{P}_b}(z)=\Omega^2 \int \text{d}^2r\sum_{ijkl} [\partial_j u_i^*(\mathbf{r})]  \eta_{ijkl} \partial_k u_l(\mathbf{r}),
\end{equation}
describes acoustic loss. In the translationally invariant case, $\alpha_{\mathcal{P}_b}$ is defined simply as a product of the inverse of the acoustic dissipation length $\alpha$ (i.e. the RHS of Eq.~(\ref{alphaP}) divided by $\mathcal{P}_b$) and acoustic energy flux $\mathcal{P}_b$. As in the optical case, we introduce the spatially averaged quantities by integrating both sides of Eq.~(\ref{dynamics.b2}) over the volume of the unit cell, arriving at
\begin{align}\label{dynamics.b3}
	\partial_z b(z) + \frac{1}{v_b} \partial_t b(z) + \frac{\langle\alpha_{\mathcal{P}_b}\rangle}{\langle\mathcal{P}_{b}\rangle} b(z) = 
	-i \Omega [a^{(1)}(z)]^* a^{(2)}(z) \frac{\langle \mathcal{Q}_b \rangle}{\langle\mathcal{P}_{b}\rangle},
\end{align}
where $v_b=\langle{\mathcal{P}_b}\rangle/\langle\mathcal{E}_{b}\rangle$ is defined similarly as for the optical fields in periodic structure. 

\subsection{Brillouin gain}

In the steady-state, Eq.~(\ref{dynamics.b3}) can be solved approximately in a similar way as we would for the regular waveguide:
\begin{align}\label{dynamics.b.3}
b(z) &\approx-i\Omega [a^{(1)}(z)]^* a^{(2)}(z) \frac{\langle \mathcal{Q}_b\rangle}{\langle \alpha_{\mathcal{P}_b}\rangle},
\end{align}
giving
\begin{align}\label{dynamics.a4}
\partial_z &a^{(1)}(z)= \omega^{(1)}\Omega a^{(1)}(z)|a^{(2)}(z)|^2 \frac{\langle \mathcal{Q}_1\rangle^* \langle \mathcal{Q}_b\rangle}{\langle\mathcal{P}^{(1)}\rangle\langle\alpha_{\mathcal{P}_b} \rangle},
\end{align}
\begin{align}\label{dynamics.a5}
\partial_z &a^{(2)}(z)= -\omega^{(2)}\Omega a^{(2)}(z)|a^{(1)}(z)|^2 \frac{\langle \mathcal{Q}_2\rangle \langle \mathcal{Q}_b\rangle^*}{\langle\mathcal{P}^{(2)}\rangle\langle\alpha_{\mathcal{P}_b} \rangle}.
\end{align}
We now approximate $\omega^{(1)}\approx\omega^{(2)}\equiv \omega$ and, follow our earlier observation that $\langle \mathcal{Q}_1\rangle=\langle \mathcal{Q}_2\rangle^*$. Furthermore, these overlap integrals can be equated to $\langle \mathcal{Q}_b\rangle$ by following the same arguments as in Refs.~\citenum{wolff2015stimulated,sipe2016hamiltonian}. We can finally define an effective Brillouin gain in an almost identical way as is done for the waveguide:
\begin{equation}\label{Bgain}
	\Gamma = 2\omega \Omega \frac{\Re[\langle \mathcal{Q}_b\rangle^* \langle \mathcal{Q}_1\rangle]}{\langle\alpha_{\mathcal{P}_b} \rangle \langle \mathcal{P}^{(1)}\rangle \langle \mathcal{P}^{(2)}\rangle}.
\end{equation}
Moreover, in the calculations we approximate $\langle\alpha_{\mathcal{P}_b} \rangle$ as $\alpha \langle{\mathcal{P}_b} \rangle$, where the \textit{spatial} dissipation rate $\alpha$ is given by
\begin{equation}
	Q_b = \frac{\Re(\tilde{\Omega})}{2\Im(\tilde{\Omega})} \rightarrow \alpha \approx \frac{\Im(\tilde{\Omega})}{v_b} = \frac{\Re(\tilde{\Omega})}{2 Q_b v_b}= \frac{\Omega}{2 Q_b v_b},
\end{equation}
giving
\begin{equation}\label{Bgain2}
\Gamma \approx 2\omega \Omega \frac{\Re[\langle \mathcal{Q}_b\rangle^* \langle \mathcal{Q}_1\rangle]}{\alpha \langle{\mathcal{P}_b} \rangle \langle \mathcal{P}^{(1)}\rangle \langle \mathcal{P}^{(2)}\rangle} = 4\omega \frac{Q_b v_b\Re[\langle \mathcal{Q}_b\rangle^* \langle \mathcal{Q}_1\rangle]}{\langle{\mathcal{P}_b} \rangle \langle \mathcal{P}^{(1)}\rangle \langle \mathcal{P}^{(2)}\rangle} = 4\omega \frac{Q_b|\langle \mathcal{Q}_1\rangle|^2}{\langle{\mathcal{E}_b} \rangle \langle \mathcal{P}^{(1)}\rangle \langle \mathcal{P}^{(2)}\rangle}.
\end{equation}

\bibliographystyle{apsrev4-1} 
\bibliography{bibliography}
\end{document}